\RequirePackage{fix-cm}
\RequirePackage[utf8]{inputenc}
\RequirePackage[T1]{fontenc}
\RequirePackage{amsmath,amssymb}

\documentclass[smallcondensed,natbib]{svjour3} % onecolumn

\usepackage{newtxtext,newtxmath}
\usepackage{textcomp}
\usepackage[caption=false]{subfig}
\usepackage{graphicx}
\usepackage[bookmarksdepth=2]{hyperref}
\usepackage{tikz}

\newcommand*{\encircle}[2][]{\tikz[baseline=(C.base)]{ % chktex 36
    \node[inner sep=0pt] (C) {\vphantom{1g}#2};
    \node[draw, circle, inner sep=0pt, yshift=1pt] 
        at (C.center) {\vphantom{1g}};}}
\newcommand*{\encirclelow}[2][]{\tikz[baseline=(C.base)]{ % chktex 36
    \node[inner sep=0pt] (C) {\vphantom{1g}#2};
    \node[draw, circle, inner sep=0pt, yshift=0pt] 
        at (C.center) {\vphantom{1g}};}}

\smartqed%
\lccode`\'=39

\begin{document}

\title{On the Dynamical Origins of Retrograde Jupiter Trojans and their Connection to High-Inclination TNOs}

\titlerunning{On the Dynamical Origins of Retrograde Jupiter Trojans}

\author{Tobias Köhne \and Konstantin Batygin}

\institute{T. Köhne (Corresponding Author)\at%
           Division of Geological and Planetary Sciences, California Institute of Technology, 1200 E California Blvd, Pasadena, CA 91125\\
           \email{tkoehne@caltech.edu}
           \and
           K. Batygin\at%
           Division of Geological and Planetary Sciences, California Institute of Technology, 1200 E California Blvd, Pasadena, CA 91125\\
           \email{kbatygin@caltech.edu}
}

\maketitle

\begin{abstract}
Over the course of the last decade, observations of highly-inclined (orbital inclination \emph{i} > 60°) Trans-Neptunian Objects (TNOs) have posed an important challenge to current models of solar system formation \citep{levison_origin_2008,nesvorny_evidence_2015}. These remarkable minor planets necessitate the presence of a distant reservoir of strongly-out-of-plane TNOs, which itself requires some dynamical production mechanism \citep{gladman_discovery_2009,gomes_observation_2015,batygin_generation_2016}.
A notable recent addition to the census of high-\emph{i} minor bodies in the solar system is the retrograde asteroid 514107 Ka'epaoka'awela, which currently occupies a 1:\textminus{}1 mean motion resonance with Jupiter at \emph{i} = 163° \citep{wiegert_retrograde_2017}.
In this work, we delineate a direct connection between retrograde Jupiter Trojans and high-\emph{i} Centaurs.
First, we back-propagate a large sample of clones of Ka'epaoka'awela for 100 Ma numerically, and demonstrate that long-term stable clones tend to decrease their inclination steadily until it concentrates between 90° and 135°, while their eccentricity and semi-major axis increase, placing many of them firmly into the trans-Neptunian domain.
Importantly, the clones show significant overlap with the synthetic high-\emph{i} Centaurs generated in Planet 9 studies \citep{batygin_planet_2019}, and hint at the existence of a relatively prominent, steady-state population of minor bodies occupying polar trans-Saturnian orbits.
Second, through direct numerical forward-modeling, we delineate the dynamical pathway through which conventional members of the Kuiper Belt's scattered disk population can become retrograde Jovian Trojan resonators in presence of Planet 9.
\end{abstract}

\section{Introduction}\label{s:intro}

Small Solar System bodies (SSSBs) like comets and asteroids have been observed for over 2000 and 200 years, respectively. Originally thought to be divine signs, SSSBs today are understood to be remnants of the planetary formation process that hold key insights into the history of our solar system. In particular, the two main characteristics of such bodies that inform the long-term evolution of the solar system are their compositions and their orbital histories. The former are of use to understand the chemical and thermal makeup of the circumstances that prevailed during the formation of the planets, and the latter yield important constraints on the dynamical evolution of the planets' orbits. While in-situ exploration of SSSBs can help shed light on their physical properties, a characterization of their dynamical evolution requires large-scale numerical simulations.

Stemming from the fact that the planet-building solar nebula is likely to have been geometrically thin, all planets and small bodies that form within it can naïvely be expected to orbit in the same direction \citep{armitage_dynamics_2011}. There have been a few exceptions to this general rule, Halley's Comet being one of the more prominent cases. Halley is a highly-eccentric comet with an aphelion beyond Pluto's orbit, but a perihelion between Mercury and Venus. It is a retrograde comet at an inclination of 162°, therefore orbiting the sun almost in the ecliptic plane, but in the opposite direction of the planets. A somewhat more recently discovered example of a retrograde SSSB is 514107 Ka'epaoka'awela \citep[2015 BZ509,][]{wiegert_retrograde_2017}. Remarkably, this object is currently entrained in a 1:\textminus{}1 mean motion resonance (MMR) with Jupiter, suggesting that it is a temporarily captured Halley-type comet. Further out in the solar system, a census of retrograde trans-Neptunian objects has also emerged within the last decade, beginning with the detection of 2008 KV42, nicknamed Drac \citep{gladman_discovery_2009}. The formation and origin of these objects remain somewhat elusive, as there is no single widely-accepted mechanism that is universally believed to generate these fascinating minor bodies \citep{volk_centaurs_2013}.

One possibility is that SSSBs in the solar system are all sourced from the Oort cloud, a nearly spherical collection of icy debris envisioned to extend outward from \( r \gtrsim 2 \cdot 10^4~\text{AU}\). The recent work of \citet{kaib_ossos_2019}, however, shows that this hypothesis falls short of explaining the prevalence of high-\emph{i} Centaurs by a factor of a few. An alternative idea is that dynamical evolution induced by the putative Planet 9 can excite large-scale variations in orbital inclinations within the distant (\( a \gtrsim 250~\text{AU}\))~Kuiper belt, and result in inward-scattering of retrograde TNOs that eventually join the Centaur population. To this end, \citet{batygin_generation_2016} and \citet{batygin_planet_2019} have shown that the presence of a ninth massive planet in the outer solar system naturally explains not only the orbital clustering of long-period Kuiper belt objects (KBOs), but also generates bodies with inclinations between 60° and 180° and perihelia smaller than 30 AU.\@ Whatever the exact generation process of \mbox{(highly-)}inclined Centaurs may be, their existence alludes to the tantalizing possibility that retrograde Jupiter Trojans and Centaurs are related. % chktex 36
Indeed, recent work has shown that retrograde MMRs might be more effective at capture than previously thought \citep{namouni_resonance_2015}.

A numerical exploration of the dynamical connection between Ka'epaoka'awela and high-\emph{i} Centaurs is the primary purpose of this work. In particular, here we report the results of two separate numerical experiments. First, in Section~\ref{s:sims_back}, clones of Ka'epaoka'awela are back-propagated to identify possible trajectories into regions in the solar system where this object could have originated. Along the way, particles that are temporarily captured onto long-lived, nearly circular and almost perpendicular orbits are identified. Second, in Section~\ref{s:sims_transfer}, we present forward-integrations of TNOs sourced from the distant scattered disk population of the Kuiper Belt, which evolve into retrograde resonance with Jupiter under the combined action of Planet 9 and the canonical giant planets.

\section{Numerical Simulations}\label{s:sims}

Numerical simulations reported in this work were performed using the \emph{Mercury6} gravitational dynamics software package \citep{chambers_hybrid_1999}. The hybrid Wisdom-Holman/Bulisch-Stoer algorithm \citep{wisdom_symplectic_1992,press_numerical_2007} is employed throughout, with an accuracy parameter set to \( \epsilon = 10^{-12}\). The initial integration time step is set at 100 days, but is automatically adapted during close encounters as needed to maintain accuracy. The ejection distance of particles leaving the simulation is set at a heliocentric distance of \( 10^4~\text{AU} \). We note that beyond this distance, the effects of passing stars as well as the galactic tide --- which we do not model here --- become appreciable. Only the four outer planets are included in the integration, and the output is recorded and analyzed at 10 ka intervals unless otherwise noted. All particles other than the Sun and the planets are set to have zero mass.

\subsection{Backpropagation of Ka'epaoka'awela}\label{s:sims_back}

In our first simulation (S1 henceforth), 6200 clones of Ka'epaoka'awela (generated using the observed orbital elements and their uncertainties, see Tab.~\ref{tab:orbelems}) are propagated back in time\footnote{We note that in chaotic dynamics, forward and backward integrations are statistically equivalent.} for 100 Ma. In an effort to accentuate regions of phase-space that correspond to diminished chaotic diffusion, long-lived particles are cloned further during the simulation. Specifically, whenever the number of test particles falls below a quarter of its initial count (due to ejection from the system or collision with the Sun or planets), leftover test particles with \( a \geq 100~\text{AU} \) are replicated\footnote{We parallelized our simulation into 62 subprocesses with 100 test particles each. The cloning is done inside these sub-simulations, such that whenever the resampling is triggered by having less than 25 of the initial 100 particles left, the leftover objects are (randomly assigned) parents for as many clones that are needed to reach 100 objects again.} and have their orbit changed slightly: The semi-major axis is altered by a fractional value \( \varepsilon \sim \mathcal{U}(-1,1) \cdot 10^{-6} \) (corresponding to a change in orbital energy comparable to the integration accuracy) to increase the effective amount of sampled potential trajectories. As a result, the final number of particles at the last time step exceeds 50000, since most clones eject from the system on a timescale considerably shorter than the integration time span.

\begin{table}
  \caption{Orbital Elements of 514107 Ka'epaoka'awela with 1\( \sigma \) uncertainty at JED 2,457,800.5 \citep{wiegert_retrograde_2017}}\label{tab:orbelems}
  \footnotesize
  \begin{tabular}{llllll}
  \hline\noalign{\smallskip}
	Semi-major &  &  & Longitude of & Argument of & Mean  \\
  axis [AU] & Eccentricity [-] & Inclination [°] & ascending node [°] & perihelion [°] & anomaly [°] \\
  \noalign{\smallskip}\hline\noalign{\smallskip}
	\( 5.140344 \) & \( 0.380661 \) & \( 163.004559 \) & \( 307.376837 \) & \( 257.449197 \) & \( 32.461694 \) \\
  \( 5.52 \cdot 10^{-5} \) & \( 5.94 \cdot 10^{-6} \) & \( 1.93 \cdot 10^{-5} \) & \( 3.41 \cdot 10^{-5} \) & \( 2.65 \cdot 10^{-4} \) & \( 4.21 \cdot 10^{-5} \) \\
  \noalign{\smallskip}\hline
	\end{tabular}
\end{table}

\begin{figure*}
	\centering
	\subfloat[10 ka\label{fig:sim_10ka}]{\includegraphics[width=\textwidth]{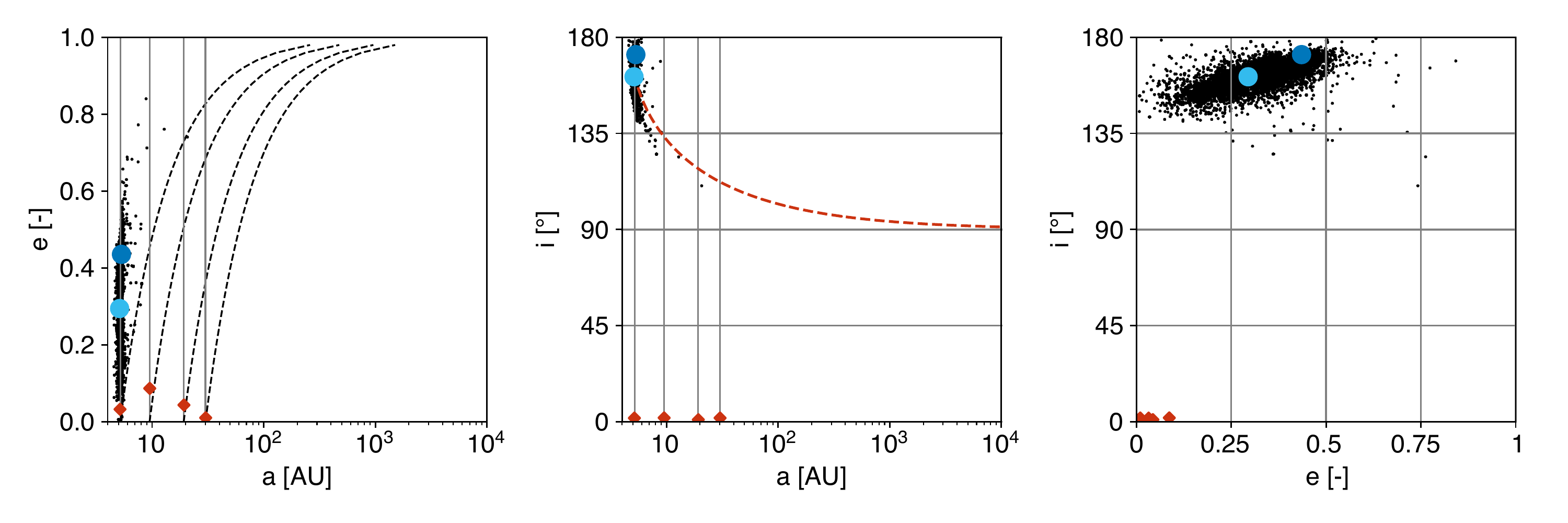}}\\
	\subfloat[1 Ma\label{fig:sim_1ma}]{\includegraphics[width=\textwidth]{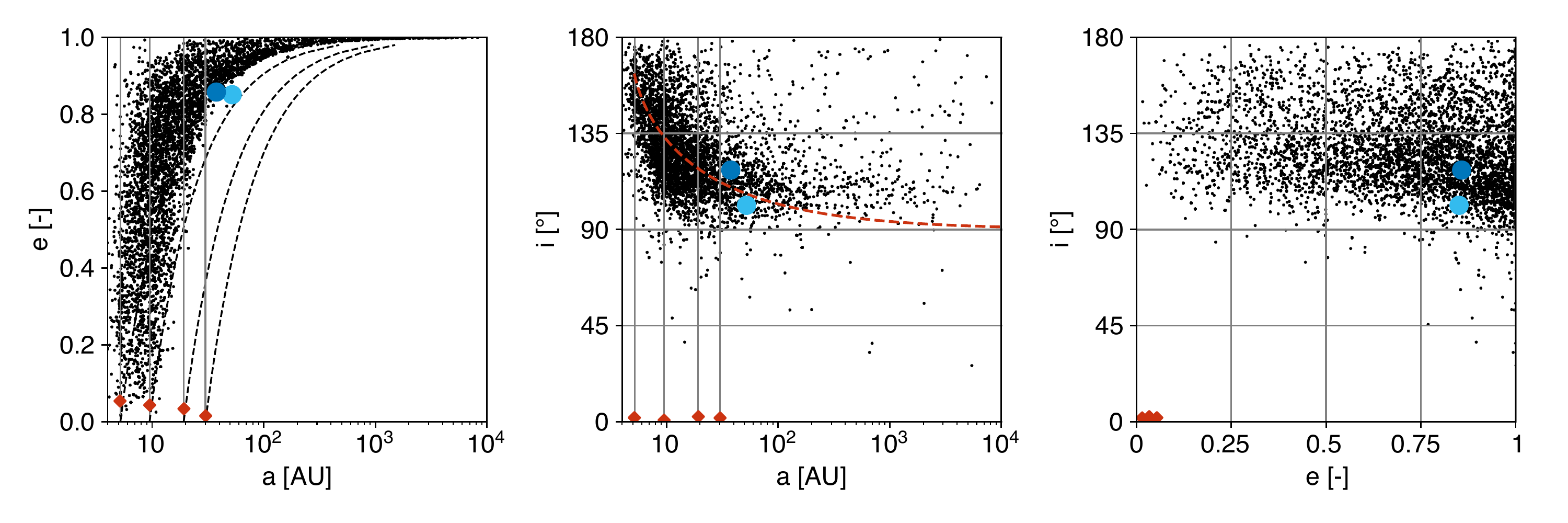}}\\
	\subfloat[100 Ma\label{fig:sim_100ma}]{\includegraphics[width=\textwidth]{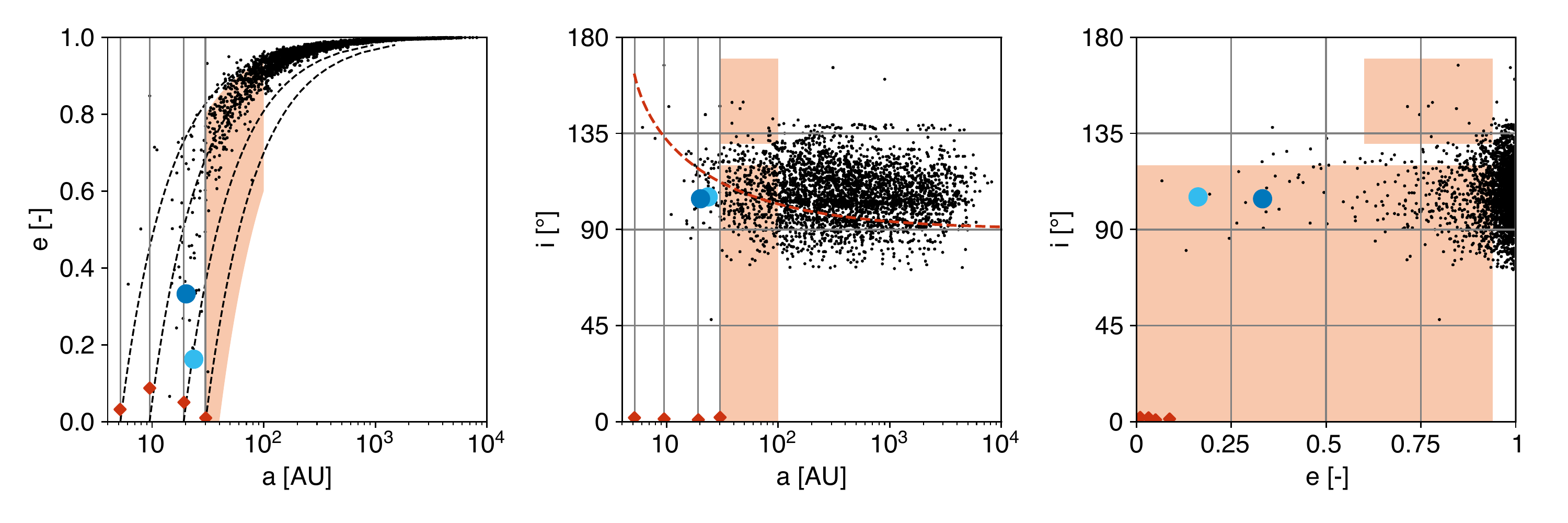}}
	\caption{Orbital elements of the test particles at (a) 10 ka, (b) 1 Ma and (c) 100 Ma, respectively. Vertical gray lines are semi-major axes of outer giant planets, which are included as red diamonds. Curved black dashed lines are lines of constant perihelion of the outer planets. The Tisserand parameter of Ka'epaoka'awela is provided for reference as the red dashed line in \emph{a}-\emph{i}-space, calculated using the orbital elements reported in \citet{wiegert_retrograde_2017} and assuming a constant \emph{e}. Test particles are shown as black dots, except for selected particles that remain in temporarily stable polar orbits and that are highlighted in Fig.~\ref{fig:sim_evolution}. In (c), the orbital domain encapsulated by synthetic populations of TNOs generated in the simulations of \citet[Fig. 1]{batygin_generation_2016} are highlighted with orange shading for reference.}\label{fig:sim_snapshots}
\end{figure*}

The first observation that can be made by examining the temporal evolution back in time for the Ka'epaoka'awela clones (Fig.~\ref{fig:sim_snapshots}) is their sensitivity to initial conditions: already after 10 ka, their orbital elements have changed drastically (the object's Lyapunov time\footnote{The Lyapunov time is estimated with the REBOUND software containing the Sun, the four outer planets and Ka'epaoka'awela. The \emph{IAS15} integrator \citep{rein_ias15_2015} is used with a time step of 100 days.} is \( \approx \) 7 ka). Notably, chaotic diffusion exhibits the clear tendency to both diminish the inclination and excite the eccentricity of the test particles.

The second snapshot at 1 Ma (Fig.~\ref{fig:sim_1ma}) shows that most clones follow the \emph{q} \( \approx \) 5--10 AU profile as they move outwards, generating an effective ``Jovian Scattered Disk''.\@ The majority of particles in the retrograde plane eject, but the most long-lived clones diffuse in non-trivial ways to inclinations between 90° and 135°. Some objects even flip to prograde motion in the presence of giant planets \citep[e.g.][]{lithwick_eccentric_2011,greenstreet_production_2012,li_survey_2019}. This leads to what essentially constitutes a selection process that filters out the short-lived particles. At 100 Ma (Fig.~\ref{fig:sim_100ma}), the simulated particles diffuse to greater perihelion distances, and the outward path along \emph{q} \( \approx \) 5--20 AU is even more prominent. \citet{namouni_interstellar_2018} find the same pattern (compare their Fig.~1) by simulating one million particles, and including galactic tides. They introduce the term ``polar corridor'', which we will adopt as well throughout this work, and define it loosely as any orbit with \(70\text{°} \leq i < 110\text{°} \).

\begin{figure*}
	\centering
	\subfloat[Cyan Particle\label{fig:sim_particle1}]{\includegraphics[width=0.49\textwidth]{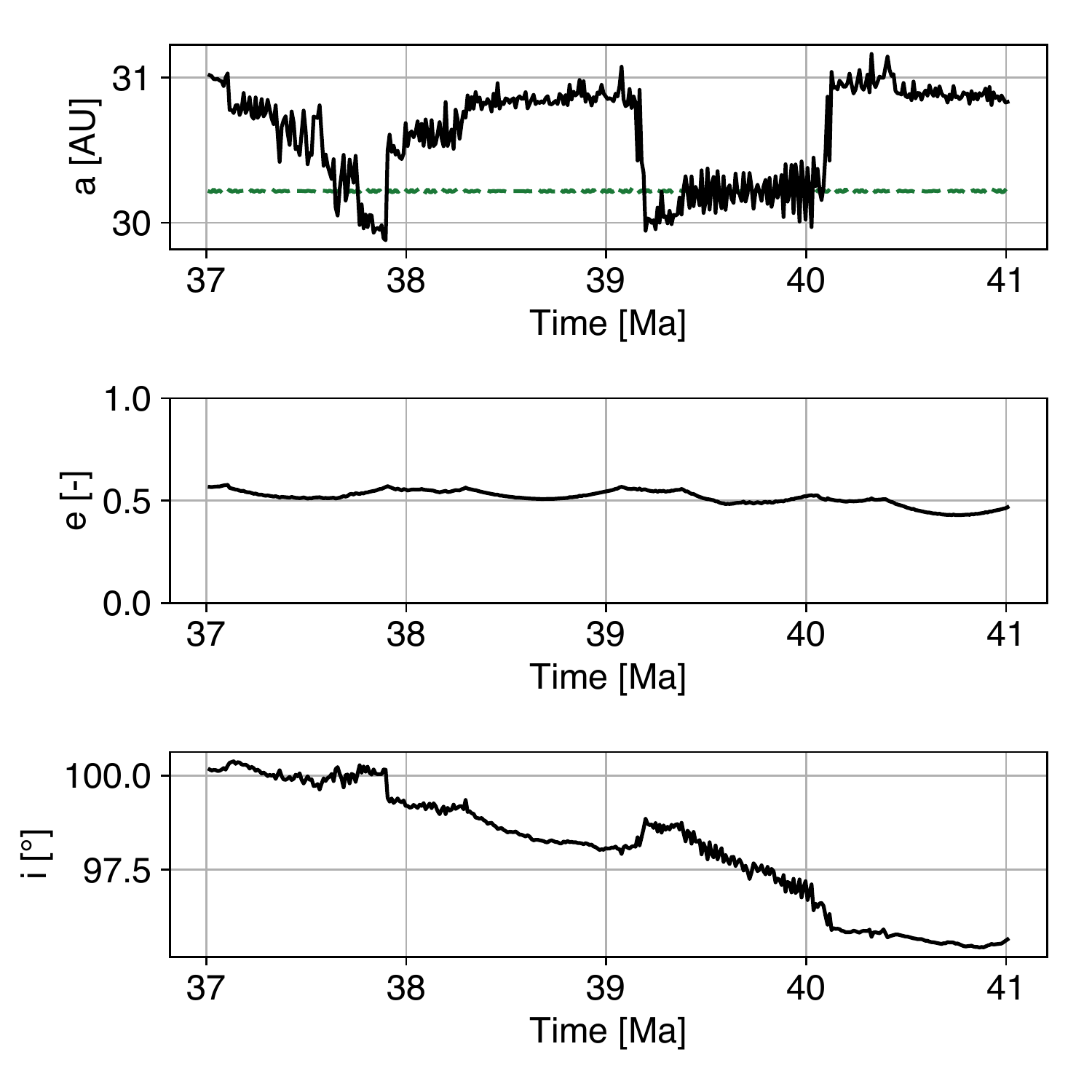}}
	\subfloat[Blue Particle\label{fig:sim_particle2}]{\includegraphics[width=0.49\textwidth]{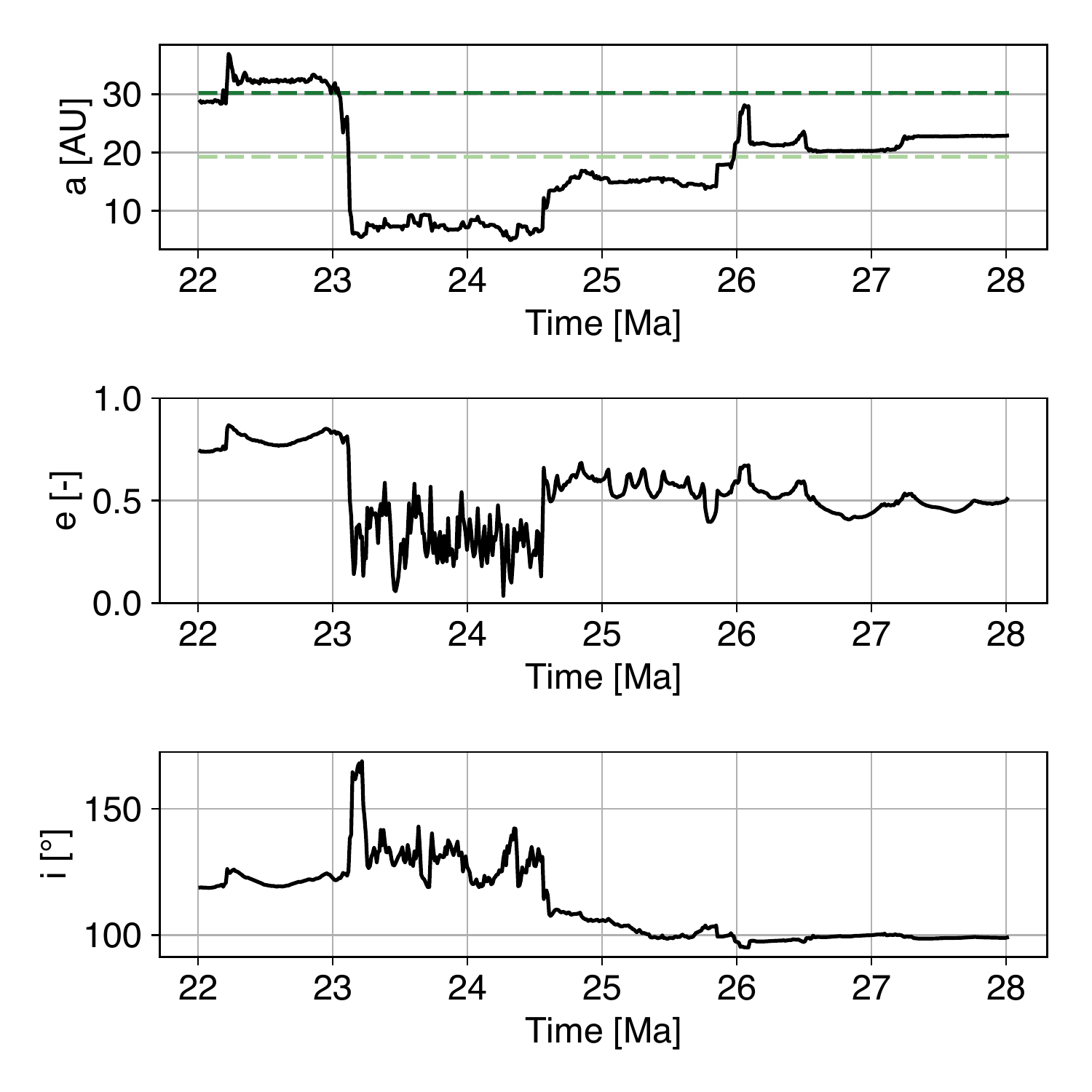}}
	\caption{Sample orbital evolution of the two test particles highlighted in Fig.~\ref{fig:sim_snapshots}. Dashed lines are Uranus (light green) and Neptune (dark green). Eccentricity and inclination of the test particles are given for reference.}\label{fig:sim_evolution}
\end{figure*}

As the simulated test particles diffuse chaotically within interplanetary space, they exhibit what appears to be temporary capture (up to several Ma) into mean motion commensurabilities at near-polar inclinations (see Fig.~\ref{fig:sim_evolution}). With the exception of the 1:1 MMR, the period ratios are not readily identifiable as first order MMRs. We note, however, that the particles in question primarily reside at high eccentricities, where the usual hierarchy of resonant strengths \citep[e.g.][]{murray_solar_2000} breaks down, meaning that temporary capture into high-order MMRs is expected.

\begin{figure*}
	\centering
	\includegraphics[width=\textwidth]{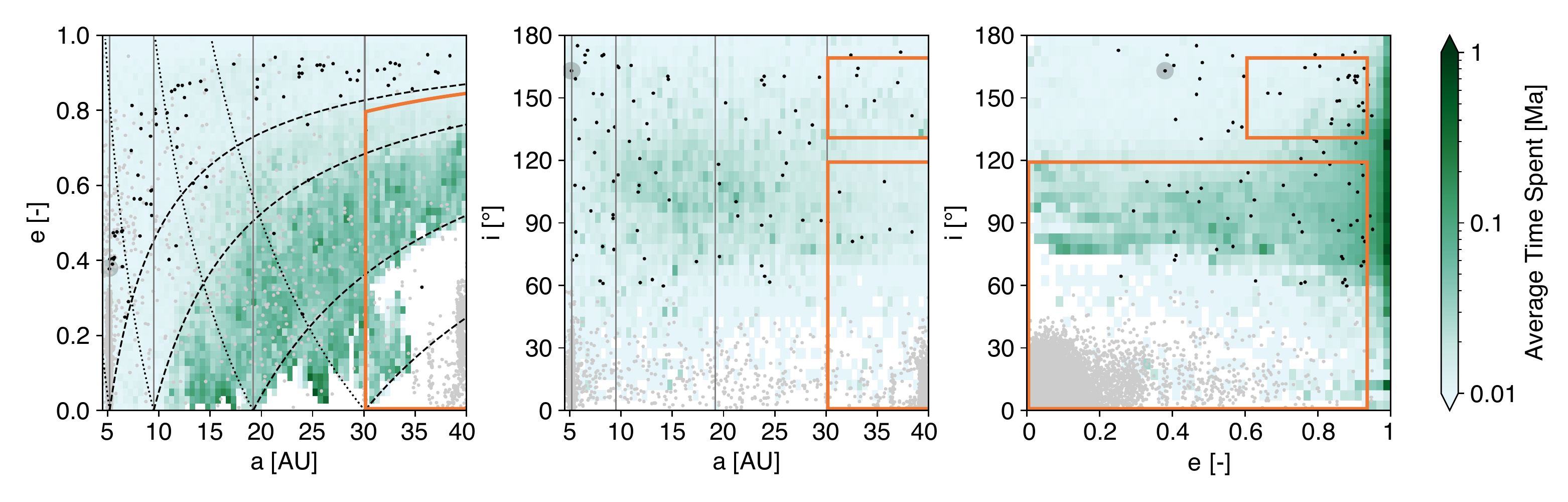}
	\caption{Average time spent of simulated S1 test particles per bin. Catalogued high-inclination (\emph{i} \( \geq \) 60°) objects are shown as black dots, other objects in gray (data by the Minor Planet Center).\@ Ka'epaoka'awela is highlighted by a gray circle. Vertical gray lines, as well as dotted and dashed curved black lines are semi-major axes, lines of constant aphelion and lines of constant perihelion of the outer giant planets, respectively. The two populations generated by \citet[Fig. 1]{batygin_generation_2016} are indicated by orange boxes.}\label{fig:mpcdensity}
\end{figure*}

Fig.~\ref{fig:mpcdensity} shows a temporal density plot of the S1 clones according to their average time spent in a particular orbit. Objects that reach a moderate-to-low-eccentricity, polar orbit have higher average lifetimes than objects that remain retrograde close to the ecliptic \citep[also see][]{gallardo_orbital_2019,li_survey_2019}. Correspondingly, these objects appear as darker shades of green in the density plot. As can also be gleaned from Fig.~\ref{fig:sim_100ma}, the dynamical pathway to generate these nearly-orthogonal orbits ensues through circularization of clones on polar orbits that reach semi-major axes approximately between Uranus and Neptune. In other words, interactions with the ice giants can reduce the test particle's eccentricity, placing them into the more stable region of \emph{a}-\emph{e}-\emph{i} space.

It is worth noting that the final state of S1 overlaps with the two synthetic TNO population groups described by \citet{batygin_generation_2016} in their investigations of the consequences of dynamical evolution induced by the proposed Planet 9. In particular, within their suite of simulations, an initially nearly-planar, prograde population of long-period (\emph{a} \( \gtrsim \) 250 AU) TNOs is excited onto high-inclination orbits via non-linear, secular interactions with Planet 9 \citep{batygin_dynamical_2017,li_secular_2018}. Subsequently, due to Neptune-scattering, a fraction of these objects is delivered inwards, eventually becoming high-inclination Centaurs. The resulting \emph{q} \( \leq \) 40 AU population occupies two regions of parameter space: One ranging between semi-major axis \( 30~\text{AU} \leq a \leq 100~\text{AU} \), perihelion \( 6~\text{AU} \leq q \leq 40~\text{AU} \), inclination \( 0\text{°} \leq i \leq 120\text{°} \) and the second one between \( 30~\text{AU} \leq a \leq 100~\text{AU} \), \( 6~\text{AU} \leq q \leq 12~\text{AU} \), \( 130\text{°} \leq i \leq 170\text{°} \). These two domains are emphasized in Figs.~\ref{fig:sim_100ma} and~\ref{fig:mpcdensity} by orange shading and boxes, respectively. Additionally, we remark that the final state of S1 also overlaps with a potential source region for high-inclination TNOs hypothesized by \citet{gladman_discovery_2009}. In their study, the authors envision it as a reservoir structure created during the final stages of planetary formation, in contrast to the dynamical generation process proposed by \citet{batygin_generation_2016}.

To validate the prevalence of the semi-stable polar orbits discussed above, we use the set of synthetic initial conditions from \citet{batygin_generation_2016} to draw 6200 test particles, and propagate them forward in time for 100 Ma. In addition to the four outer planets (and in contrast to S1), an effective J2 moment is added to the Sun to approximate the influence of the inner planets\footnote{Although the initial conditions for the TNO population have been taken from a simulation with Planet 9, the particles have been propagated forward in time without Planet 9, as we do not expect the planet to have a significant influence on the test particles' trajectories once they begin their journey to the inner planets. The observation, that these high-inclination TNOs exist in the first place, is independent of their assumed method of generation: We merely show that their population could explain the presence of polar SSSBs co-orbital to the giant planets.}. The Sun's radius is increased to 4 AU to capture test particles that are scattered into the inner solar system and remove them. No cloning of particles is performed. As in S1, the temporarily stable, polar orbits with semi-major axes approximately between Uranus and Neptune are naturally produced. Specifically, we find test particles on trajectories very similar to Fig.~\ref{fig:sim_evolution} (but do not show them here to avoid redundancy). The number of test particles on long-lived orbits is at least 1 in 1000 by the end of the simulation when compared to the initial particle count, suggesting that the number density of such bodies in the solar system may be non-negligible.

\subsection{TNO Transfer Rates into Retrograde Resonances}\label{s:sims_transfer}

\begin{figure*}
	\centering
	\includegraphics[width=\textwidth]{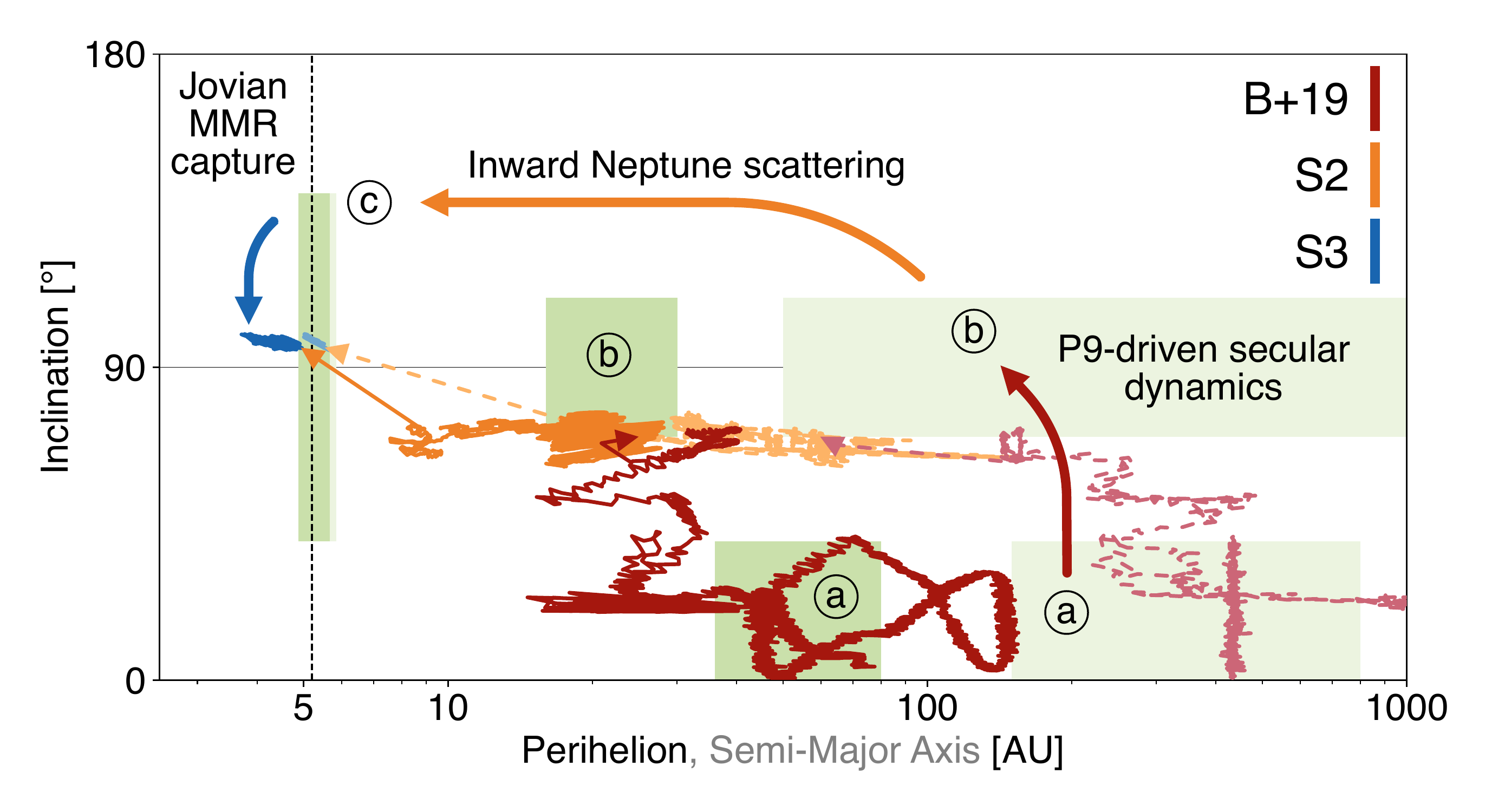}
	\caption{Schematic representation of the multi-stage simulations described in Section~\ref{s:sims_transfer}. Shown is an example trajectory of a particle that started out in low-inclination, prograde orbit beyond Neptune, evolved through the polar corridor into Jupiter's vicinity, and then entered a 1:\textminus{}1 retrograde resonance (the tip of the arrow points at the last time step). The colors correspond to the different simulations, with {B+19} referring to the numerical experiments of \citet{batygin_planet_2019}. The dark, solid lines are the particles' orbit in \emph{a}-\emph{i} space, whereas the fainter, dashed lines denote \emph{q}-\emph{i} coordinates. The green boxes represent approximate regions of interest (darker coloring in terms of \emph{a}, fainter shading for \emph{q}): \encirclelow{a} the initial conditions of {B+19}, \encircle{b} the orbital parameter space of particles simulated by {B+19} that have entered the polar corridor and are the basis for S2, and \encirclelow{c} the orbital parameter space for particles in S2 that reached a semi-major axis close to Jupiter and are the initial conditions for S3. The transfer mechanisms of each simulation are given next to the respective schematic arrows.}\label{fig:stepvis}
\end{figure*}

While S1 elucidates the potential regions of parameter space from which Ka'epaoka'awela could have originated, it falls short of demonstrating the actual generation pathway of retrograde Trojans from any specific source region. Ideally, one would start out with the same initial conditions as the aforementioned \citet{batygin_planet_2019} simulations (or alternatively, any model of source regions to investigate, e.g.\@ the Oort cloud), increase the particle count, and continue simulating objects until (by chance) creating objects that reach a retrograde resonance with Jupiter.\@ Naïvely, the probability of generating a retrograde Trojan population would then just be their fraction of the integrated particles, and one could compare this with the generation probability of a prograde Trojan, or any other asteroid family. However, this would require an unreasonable computational effort, so we therefore choose an alternative, multi-stage approach. First, we extract trajectories simulated by \citet{batygin_planet_2019} that at some point in their evolution (but at least after two billion years into their integration) reach an orbit of semi-major axis \( a < 1000~\text{AU} \), perihelion \( q < 30.1~\text{AU} \) and inclination \( 70\text{°} \leq i \leq 110\text{°} \). These are the particles that we define to have entered the polar corridor, and overlap with the observed high-\emph{i} SSSB population. We find that for the \( 10^4 \) test particles present, 119 meet these criteria. In our second set of simulations (S2), we propagate those particles forward in time to determine their likelihood to reach any of the outer planets. Lastly, for the particles that come close to Jupiter in terms of orbital periods, we assess their capture probability in our third and final simulation (S3). A schematic representation of the process is depicted in Fig.~\ref{fig:stepvis}.

S2 therefore aims to characterize the transfer rate between TNOs that entered the polar corridor and potential Jupiter capture candidates (Fig.~\ref{fig:stepvis}, orange lines). The integration includes the four outer planets as well as Planet 9\footnote{We performed the same simulation without Planet 9 and reach comparable results. This is expected since the particles observed are far enough away from Planet 9 to not be significantly affected. We therefore omit a more detailed presentation in order to avoid repetition of the conclusions.}. An effective J2 moment is added to the Sun (with its radius is increased to 4 AU) to approximate the influence of the inner planets and remove particles that enter the terrestrial region of the solar system. The simulation is initialized with 2000 particles, sampled from the 119 TNOs that entered the polar corridor (Fig.~\ref{fig:stepvis}, box \encircle{b}), with a small perturbation added to the semi-major axis (on the order of magnitude of the orbital energy integration error, see S1), and randomized mean anomaly. Whenever a particle is removed from the simulation (either through collision, ejection, or by increasing the perihelion to \( q > 30.1~\text{AU} \)), a new one is added again, sampled the same way as the initial particles\footnote{In this case, the positions of the planets are therefore different to what the initial test particle population experienced. This is not a problem, however, for two reasons: First, the 119 original particles were extracted at different times (i.e., positions of planets) already, but most importantly, the entering of the polar corridor is, to an excellent approximation, independent of the positions of the planets \citep{batygin_dynamical_2017,li_secular_2018}.}. The integration is run for one billion years, over the course of which 51135 trajectories are generated.

To determine the transfer rates from the initial population to the four outer planets, we record all particles that reach a semi-major axis that is within three Hill radii of any planet's semi-major axis. We find 432, 186, 53, and 2 particles that approach Neptune, Uranus, Saturn, and Jupiter (Fig.~\ref{fig:stepvis}, box \encirclelow{c}), respectively, consistent with the standard picture of chaotic diffusion of TNOs into the intra-Neptunian region of the solar system\footnote{We note that this is most likely an underestimate, since from a code framework perspective, we are restricted to analyzing the orbits at the output timesteps of 10 ka. We are therefore missing all particles that enter and leave a planet's influence in between the snapshots.}. Of the two particles that reached Jupiter, one reached the planet at a prograde inclination (\( i \approx 41\text{°} \)), and one at a polar retrograde inclination (\( i \approx 93\text{°} \)). The exact orbital configuration is important at this stage, as the difference between collision, ejection and capture due to the impending contact is sensitive to variations in the relative position on a kilometer scale. As such, it is not important that the specific two particles we found near Jupiter did not immediately get captured into a stable, resonating orbit. Instead, the capture probability must be estimated with a final set of simulations.

In S3 (Fig.~\ref{fig:stepvis}, blue lines), we approximate this estimate by cloning the retrograde capture candidate particle near Jupiter at the last timestep covered by S2. The clones only differ in semi-major axis by a fractional value \( \varepsilon \sim \mathcal{U}(-1,1) \cdot 10^{-9} \). S3 still contains all outer planets as well as the inner planet's effective J2 moment added to the Sun, but we omit Planet 9, as it has a negligible influence at this point, and decrease the Sun's radius to 1.7 AU in order not to remove prematurely any particles that have \( q \leq 4~\text{AU}\). Because we are interested in a fine-scale assessment of orbital resonances with Jupiter, we decrease the output time step to one year (the initial integration time step is still 100 days), and monitor the 1:\textminus{}1 libration angle. The propagation is run for 10 ka, but for particles caught into a resonant cycle of libration, we continue their integration until the resonance is interrupted to characterize the resonance lifetimes. Of 350 injected clones, we find that 14 enter into a 1:\textminus{}1 libration for at least 500 years (median 4500 years), corresponding to a capture rate of 4\%\footnote{We find a similar capture rate into the 1:1 resonance for the clones of the prograde particle.}.

Our piecewise continuous simulations therefore demonstrate a viable generation pathway for Ka'epaoka'awela-like objects: the simulations of \citet{batygin_planet_2019} show the injection of 119 out of \( 10^4 \) particles into the polar corridor over a timescale of two billion years; our S2 simulation shows how one of these 119 particles (and 51135 clones) attains a semi-major axis close to that of Jupiter over a timescale of one billion years; and finally, our S3 simulations produce Jovian resonance capture for 14 out of 350 clones of this object over 10 ka. Indeed, if the pre-Jovian capture dynamics of Ka'epaoka'awela followed a pathway that is similar to the one envisioned herein, it must have originated from a markedly numerous source population.

We want to point out at this stage that we by no means imply that the pathway from initially prograde, nearly-planar TNO, through the polar corridor, into a Jupiter resonance is the only possible (or even most likely) generation mechanism for retrograde Jovian Trojans, nor do we see our results as another hint at the existence of Planet 9. To this end, we purposely avoid concatenating the particle counts reported above into a single transfer rate or even mass flux because of the additional assumptions that would need to be made to combine our simulation results. In fact, the key takeaway from this section should only be that starting from a population of high-inclination Centaurs (which is observationally confirmed), chaotic diffusion is a sufficient explanation for Ka'epaoka'awela-like objects. The particles entering the polar corridor, however, can be sourced from different regions of the solar system. Apart from the excitation of TNOs by Planet 9, \citet{kaib_ossos_2019}, for example, have shown that the Oort cloud is another plausible explanation. In their analysis, they can reproduce the observed populations for high-inclination, scattering particles to within a factor of a few, and are able to match it even better by assuming a (reasonably) denser birth cluster for the solar system.

\section{Conclusion}\label{s:conclusion}

Based on (1) the overlap of the orbital element space populated by particles simulated in S1 and the observed high-\emph{i} Centaur population, and (2) our piecewise continuous simulation of inward-scattered TNOs entering a retrograde 1:\textminus{}1 resonance with Jupiter (S2, S3), we confirm the naïve suspicion that Ka'epaoka'awela could have originated as a trans-Jovian minor body \citep[in contrast to the interpretation of][]{namouni_interstellar_2018}. We note, however, that more work is required to accurately quantify the Centaur transfer rate between the proposed high-\emph{i} distant source and the Jovian co-orbital state. Nevertheless, if this scenario is correct, the surface of Ka'epaoka'awela could bare the signs of substantial interstellar irradiation, since high-\emph{i} Centaurs are believed to be sourced from far beyond the heliopause \citep{hudson_laboratory_2008}. This could motivate a more detailed study of its physical properties, but a full analysis is well beyond the scope of this work.

The simulation of the trajectory of clones of Ka'epaoka'awela additionally indicates that polar orbits can experience increased longevity in the vicinity of Uranus and Neptune \citep[compare][]{gallardo_orbital_2019}. While we have not identified libration of specific resonant angles potentially exhibited by the simulated test particles, the quantized ratios of \emph{a} between the planets and the diffusing particles, as well as their temporarily enhanced dynamical stability, likely point to the stochastic operation of the resonant phase-protection mechanism.

Currently, strong observational biases act against the detection of high-\emph{i} Centaurs such as the ones simulated in this work. Our numerical experiments, however, hint at a pronounced existence of minor bodies with \emph{e} \( \lesssim \) 0.5, \emph{i} \( \approx \) 90° and \( a_\text{Uranus} \lesssim a \lesssim a_\text{Neptune} \). Correspondingly, a sufficiently deep, all-sky survey may have a good chance of unveiling this inferred population of such objects observationally. To this end, it is worth noting that the Large Synoptic Survey Telescope is expected to begin operations in 2023 and increase the catalog of SSSBs by a factor of 10--100 \citep{jones_asteroid_2015}. The simulations performed in this work therefore suggest that as the census of highly-inclined Centaurs comes into view, the dynamical link between retrograde Jupiter Trojans and TNOs will also come into sharper focus.

\begin{acknowledgements}
We thank Nathan A. Kaib and a second anonymous reviewer for their valuable comments and suggestions for improving this manuscript. This research has made use of data and/or services provided by the International Astronomical Union's Minor Planet Center. Simulations in this paper made use of the REBOUND code which is freely available at https://github.com/hannorein/rebound. KB gratefully acknowledges the David and Lucile Packard Foundation and the Alfred P. Sloan Foundation for their generous support.
\end{acknowledgements}

\bibliographystyle{spbasic}
\bibliography{Manuscript_arxiv}

\end{document}